\documentclass[%
 aip,
 amsmath,amssymb,
preprint,%
]{revtex4-1}

\usepackage{graphicx}
\usepackage{dcolumn}
\usepackage{bm}

\usepackage[utf8]{inputenc}
\usepackage[T1]{fontenc}
\usepackage{mathptmx}
\usepackage{etoolbox}
\usepackage{mhchem}
\usepackage{graphics}

\makeatletter
\def\@email#1#2{%
 \endgroup
 \patchcmd{\titleblock@produce}
  {\frontmatter@RRAPformat}
  {\frontmatter@RRAPformat{\produce@RRAP{*#1\href{mailto:#2}{#2}}}\frontmatter@RRAPformat}
  {}{}
}%
\makeatother
\begin{document}

\preprint{}

\title[]{Microscopic Mechanism of the Thermal Amorphization of ZIF-4 and Melting of ZIF-zni Revealed via Molecular Dynamics and Machine Learning Techniques}

\author{Emilio Mendez}
\author{Rocio Semino}%
 \email{rocio.semino@sorbonne-universite.fr}
\affiliation{ 
Sorbonne Université, CNRS, Physico-chimie des Electrolytes et Nanosystèmes Interfaciaux, PHENIX, F-75005 Paris, France
}%

\date{\today}%

\begin{abstract}
We investigate the microscopic mechanism of the thermally induced ambient pressure ordered--disordered phase transitions of two zeolitic imidazolate frameworks of formula \ce{Zn(C3H3N2)2}: a porous (ZIF-4) and a dense, non-porous (ZIF-zni) polymorph \textit{via} a combination of data science and computer simulation approaches. Molecular dynamics simulations are carried out at the atomistic level through the nb-ZIF-FF force field that incorporates ligand--metal reactivity and relies on dummy atoms to reproduce the correct tetrahedral topology around \ce{Zn^2+} centres. The force field is capable of reproducing the structure of ZIF-4, ZIF-zni and the amorphous (ZIF$\_$a) and liquid (ZIF$\_$liq) phases that respectively result when these crystalline materials are heated. Symmetry functions computed over a database of structures of the four phases, are used as inputs to train a neural network that predicts the probabilities of belonging to each of the four phases at the local \ce{Zn^2+} level with 90\% accuracy. We apply this methodology to follow the time-evolution of the amorphization of ZIF-4 and the melting of ZIF-zni along a series of molecular dynamics trajectories. We first computed the transition temperature and determined associated thermodynamic state functions. Subsequently, we studied the mechanisms. Both processes consist of two steps: (i) for ZIF-4, a low-density amorphous phase is first formed, followed by the final ZIF$\_$a phase while (ii) for ZIF-zni, a ZIF$\_$a-like phase precedes the formation of the liquid phase. These processes involve connectivity changes in the first neighbour ligands around the central \ce{Zn^{2+}} cations. We find that the amorphization of ZIF-4 is a non-isotropic processes and we trace back the origins of this anisotropic behaviour to density and lability of coordination bonds.  

\end{abstract}

\maketitle

\section{\label{sec:intro}INTRODUCTION}

Amorphous Metal-Organic Frameworks (MOFs) have been long known, but only very recently they have attracted attention of the research community.~\cite{Horike2020} Indeed, since amorphous MOFs may conserve building blocks and practically all connectivity from their crystalline counterparts, they combine attractive properties such as intrinsic porosity and high surface areas~\cite{Thornton2016} of crystalline phases with  mechanical robustness and the presence of multiple defects that can act as catalytic centres typical of amorphous phases.~\cite{Ma2022} Amorphous MOFs can be synthesised as such, but they are mostly obtained from their parent crystalline structures by exerting an external stimulus on them.~\cite{Fonseca2021} Since these ordered--disordered transitions are reversible, guests within the pores of the crystalline phases can become trapped when the structure becomes amorphous to be later released by forcing the MOF to return to its crystalline state. This principle makes these materials attractive for important industrial and environmental applications including water purification,~\cite{Zhang2019} drug delivery~\cite{OrellanaTavra2016,OrellanaTavra2020}, capture of radioactive species~\cite{Sava2011,Chapman2011} and catalysis~\cite{Yu2017,Duan2019} among others. Moreover, it is easier to adequately shape amorphous materials for applications (for example as pellets, extrudates or sprays) without compromising their porosity or chemical properties.~\cite{BazerBachi2014} 

Zeolitic Imidazolate Frameworks (ZIFs) conform an exceptionally stable family of MOFs with potential applications to many societal challenging processes. These MOFs have the particularity that their distribution of metal--ligand--metal angles is analogous that of Si--O--Si angles in zeolites, but since their bonds are longer, their porosities are larger. This offers some advantages in terms of potential guest-host-based applications, but it also gives ZIFs more flexibility (i.e. their elastic moduli are at least one order of magnitude lower than those of zeolites),~\cite{Tan2010} as coordination bonds are weaker than covalent bonds. In addition, amorphization of ZIFs occurs at milder conditions (for example, lower temperatures) than for their zeolite analogues,~\cite{Greaves2003,Peral2006} as deforming the \ce{metal-ligand4} tetrahedron involves reorganising much weaker bonds than it does for the \ce{SiO4} case.         

Many ZIFs exhibit ordered--disordered phase transitions. These may be induced by changes in temperature,~\cite{Bennett2010,Bennett2011} pressure,~\cite{Bennett2011_3} mechanical grinding,~\cite{Bennett2011_2} interaction with X-rays,~\cite{Widmer2019_2} or even eliminating water from their structures.~\cite{Wei2023} Heat-induced amorphization was observed in a number of MOFs,~\cite{Masciocchi2001,Ohara2009,Ma2022,LenAlcaide2023} in this work we will concentrate our efforts on ZIF-4. This MOF is well-known for its applications to separating alkenes from alkanes among other gas mixtures~\cite{Hartmann2015} and its synthesis has been scaled-up.~\cite{Hovestadt2017} It is one of the many existing polymorphs of chemical formula \ce{Zn(C3H3N2)} and it exhibits a \textit{cag} topology with connected cages of a diameter of 4.9 \AA.~\cite{Park2006} ZIF-4 has a complex phase diagram, consisting of a series of amorphous and crystalline phases.~\cite{Widmer2019} At ambient pressure, the following phase transitions have been experimentally detected: 

\begin{center}
    ZIF-4 $\xrightarrow{\text{      1      }}$ ZIF$\_$a $\xrightarrow{\text{       2       }}$ ZIF-zni $\xrightarrow{\text{       3       }}$ ZIF$\_$liq
\end{center}
        
\emph{Transition 1} is an \textit{amorphization} phase transition from the crystalline porous ZIF-4 to ZIF$\_$a, an amorphous phase that has a continuous random network structure similar to amorphous silica.~\cite{Bennett2010} This transition occurs at T$_1$=589 K.~\cite{Bennett2016} Further heating leads to \emph{transition 2}: a recrystallization of ZIF$\_$a into the crystalline dense ZIF-zni solid, which happens at T$_2$=773 K.
It is interesting to note that other \ce{Zn(C3H3N2)} polymorphs, including ZIF-1 (\textit{crb} topology), ZIF-3 (\textit{dft} topology) and ZIF-6 (\textit{gis} topology) also yield the same amorphous phase upon heating that subsequently crystallises into ZIF-zni.~\cite{Bennett2011} Finally, \emph{transition 3} involves the melting of ZIF-zni at T$_3$=863 K to give a liquid MOF (ZIF$\_$liq).~\cite{Beake2013} Bennett and coworkers have shown that heating ZIF-4 first leads to the formation of a low density phase before reaching the high density amorphous phase that transforms into ZIF-zni upon further increasing the temperature.~\cite{Bennett2015}

Neutron total diffraction data revealed that the Zn-centred tetrahedron remains quite rigid during the amorphization, although an out-of-the-plane ligand motion can act slightly reducing the total connectivity of the amorphous phase with respect to the full connectivity of ZIF-4.~\cite{Beake2013} High field $^{13}$C and $^{15}$N NMR measurements can help differentiate ZIF-4 from ZIF-zni and show that there is an important structural similarity between ZIF$\_$a and both crystalline phases, albeit a signal broadening for the amorphous material. This confirms that the amorphous phases obtained by thermal annealing starting from ZIF-4 or ZIF-zni are identical.~\cite{Baxter2015} \textit{In situ} far infra-red spectroscopy proved that the ligand does not significantly strain during the amorphization process and that collective modes involving deformation of the \ce{ZnN4} tetrahedron are the main contributors to the amorphization.~\cite{Ryder2017} From a computational standpoint, Gaillac and coworkers have studied the melting mechanism of ZIF-4 \textit{via ab initio} molecular dynamics simulations.~\cite{Gaillac2017,Gaillac2018} In these works, ZIF-4 is considered as a starting point, since treating representative sections of ZIF$\_$a and ZIF-zni would yield too large systems to subject them to \textit{ab initio} molecular dynamics while spanning reasonable timescales to get correct simulation averages. Very high temperatures, of the order of 1000 K or more, are explored in order to get enough statistics, so the intermediate states between ZIF-4 and ZIF$\_$liq are ignored. The authors reach a good agreement between the structural characteristics of their model and the experimental PDFs, they find that under-coordinated Zn centres act as "seeds" for the melting process to occur and they propose a molecular mechanism for the melting involving first-neighbours exchanges. Despite this wealth of information, many questions remain unanswered, including: what is the mechanism of the amorphization of ZIF-4? What is the mechanism of the melting of ZIF-zni? What happens beyond first-neighbours distances in these transformations? In order to answer these questions we need larger simulation cells, thus, it is necessary to rely on a computational model where electronic degrees of freedom are averaged following the Born-Oppenheimer approximation. However, including reactivity, in particular, metal--ligand reactivity, is essential to model these kinds of ordered--disordered phase transitions. A popular reactive force field, ReaxFF, has been tested for this task, but the modelling community has not yet reached a consensus on whether this force field is adequate to model amorphous ZIFs.\cite{Yang2018,Castel2022}

In this contribution, we rely on nb-ZIF-FF (non-bonded ZIF-FF)~\cite{Balestra2022} to study the ordered--disordered \emph{transitions 1} and \emph{3} cited above. This force field, originally developed by Balestra and Semino with the purpose of modelling the self-assembly of ZIFs, features metal--ligand reactivity by means of Morse potentials to treat coordination bonds. We found that it captures structural, mechanical and thermodynamic properties of the amorphous and liquid phases as well as of the crystalline ones. We circumvent the inherent difficulty of differentiating multiple ordered and disordered phases through the training of a neural network sorting algorithm that identifies the correct phase at the local, atomic level with high accuracy. This data science augmented molecular dynamics approach allows us to reach molecular detail in the study of the mechanisms of these ordered--disordered transformations and to answer the above raised questions. We observe the formation of a phase recognised as liquid-like prior to the formation of ZIF$\_$a upon heating ZIF-4 at a temperature T>T$_1$, which can be associated with experimental observations.~\cite{Bennett2015} For ZIF-zni, we found the opposite: the density decreases monotonically, going from an amorphous-classified state into the liquid. Both processes occur in an anisotropic fashion, which can be correlated to the local density and labilities of coordination bonds within the materials.  

This work is structured as follows. Sec. \ref{sec:methods} details the systems of study, the algorithm applied to generate disordered structures, the development of our neural network sorting algorithm and the molecular dynamics simulations details. We then present and discuss our results in Sec.~\ref{sec:res} and summarise them in Sec.~\ref{Sec:Conc}.

\section{\label{sec:methods}METHODS}

\subsection{\label{sec:sys}Systems of Study}

We focus on two crystalline MOFs within the \ce{Zn(C3H3N2)2} series of polymorphs:~\cite{Park2006} ZIF-4 and ZIF-zni, a porous and a dense phase respectively. We analyse the phase transitions that transform ZIF-4 into ZIF$\_$a and ZIF-zni into a melt: ZIF$\_$liq (\emph{transition 1} and \emph{transition 3}, see Sec. \ref{sec:intro}).~\cite{Widmer2019} These two are ordered--disordered transitions, so we can adequately sample them by just increasing the temperature (to increase the kinetic energy) of the materials. Studying \emph{transition 2} is a much more complicated problem, since it would involve using more sophisticated enhanced sampling simulation methods, to correctly sample the collection of rare events that lead from a disordered to an ordered phase. 

\begin{figure}
\vspace{-1cm}
\includegraphics[width=0.5\textwidth]{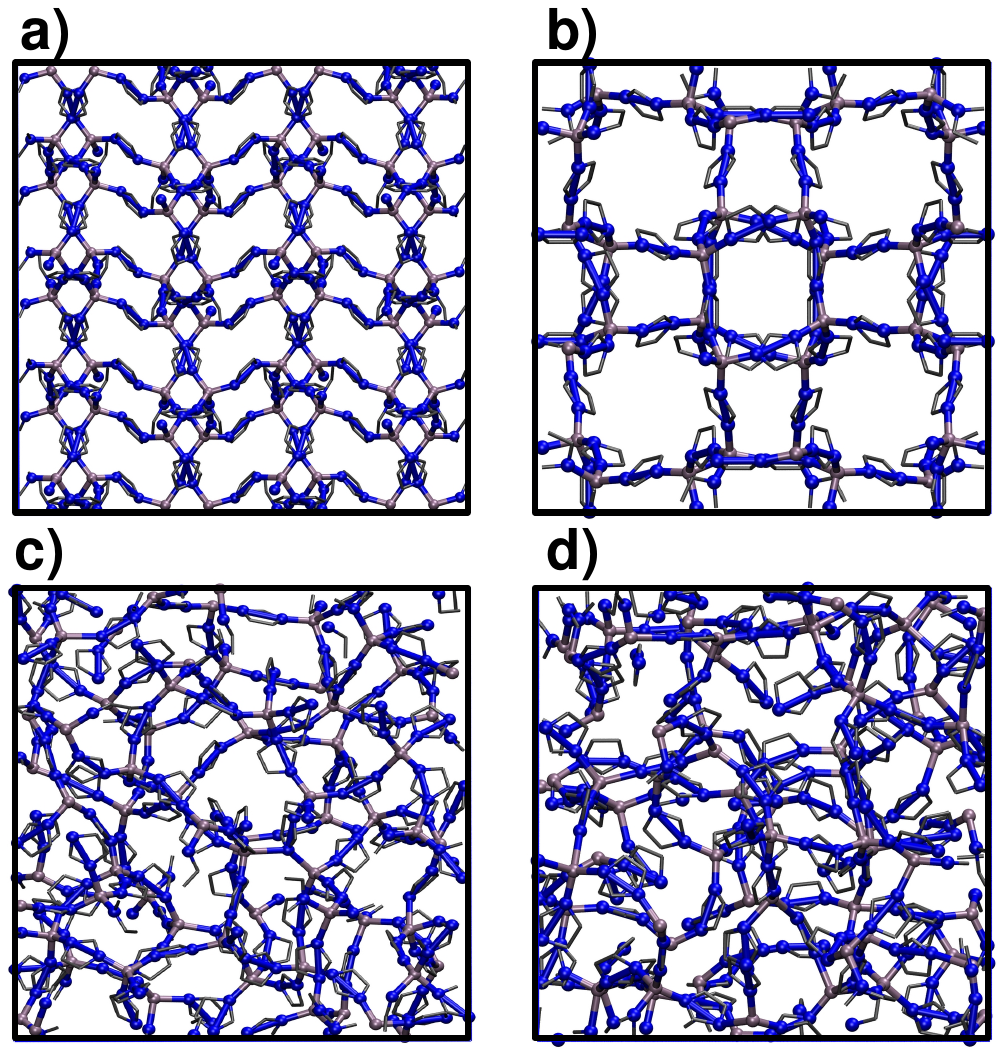}
\vspace{-3.5cm}
\caption{\label{fig:mofs} Structure of (a) ZIF-4, (b) ZIF-zni, (c) ZIF$\_$a, (d) ZIF$\_$liq. Color code: Zn (silver), N (blue), C (grey). For clarity purposes, H atoms are ignored and the snapshots have been scaled up to the same size.}
\end{figure}

Fig.~\ref{fig:mofs} illustrates the structure of the four studied systems. ZIF-4 and ZIF-zni initial structures were obtained from the Cambridge Structural Database.\cite{Groom2016} ZIF$\_$a configurations were generated by simulated annealing of these two crystalline phases \textit{via} the procedure that we detail below, in Sec. ~\ref{sec:amorphgen}. ZIF$\_$liq is obtained by heating ZIF-zni or ZIF-4 to T = 700 K. All materials were modelled at the atomistic level through the nb-ZIF-FF force field.\cite{Balestra2022} This force field partially includes reactivity, by allowing the metal--ligand bond to break and form throughout classical molecular dynamics simulations. This is the only kind of reactivity needed to model amorphization of ZIF-4, since the integrity of the ligands along the process was confirmed experimentally.\cite{Ryder2017} The metal--ligand bond is modelled through a Morse potential, which yields zero pairwise forces at long metal--ligand distances. The correct tetrahedral local symmetry around \ce{Zn^{2+}} cations is enforced by distributing part of the cation's charge into the four vertices of a flexible tetrahedron that holds it in its centre (cationic dummy atom model). This force field has been carefully validated for reproducing structural and mechanical properties of a series of ZIFs, including ZIF-4 and ZIF-zni. For more details concerning the nb-ZIF-FF force field, the reader is referred to Ref.~\onlinecite{Balestra2022}.

\subsection{\label{sec:amorphgen}Generation of the Amorphous Structures}

To generate ZIF$\_$a, we started either from ZIF-4 or from ZIF-zni. The crystalline structures were subjected to a temperature ramp through which the material is heated from 300 to 900 K, then a short constant T=900 K run is performed and finally a second temperature ramp brings the system back to 300 K. We tested ramps of 1 K ps$^{-1}$ and 4.8 K ps$^{-1}$, which comply with the less-than-5 K ps$^{-1}$ rule proposed by Castel and Coudert to avoid damaging the full coordination of the \ce{Zn^{2+}} centres in the crystalline form.\cite{Castel2022} We note that these ramps are orders of magnitude higher than those experimentally achievable (on the order of 5-100 K min$^{-1}$)\cite{Bennett2015}. We compared radial distribution functions (RDFs) of the amorphous materials generated with these two temperature ramp values and found no significant differences.
Constraints were added on the 900 K constant temperature simulation so that the final materials have approximately a cubic shape.

To sample the diversity of glass configurations, we generated three independent runs starting from ZIF-4 and three starting from ZIF-zni, that differ in the total time span of the 900 K constant temperature section: 250, 500 or 750 ps. The calculated properties of the glass correspond to an average of the six independent structures obtained after the annealing. During the heating ramp, the volume of the system suffers from an abrupt change that can be distinguished from the expected fluctuations, signalling the phase transition. Even though at the end of the simulation experiment the temperature is brought back to its initial value of 300 K, the final volume differs from the initial one, thus demonstrating that the phase transition has taken place and has not reverted. 

ZIF$\_$a configurations generated from ZIF-4 or from ZIF-zni are indistinguishable, and they describe quite well the experimentally characterised amorphous material in terms of density, typical neighbour distances and angles and of its bulk modulus, as shown in Tab.~\ref{tab:amorphous}. 
The bulk modulus $K$ was calculated from the volume fluctuations in the NPT ensemble:
\begin{equation}
    \langle(\Delta V)^2\rangle = \frac{k_b T}{K}
\end{equation}

\begin{table}
    \caption{\label{tab:amorphous}}Density, bulk modulus, Zn--Zn and Zn--N average neighbours distances and N--Zn--N average angle computed \textit{via} nb-ZIF-FF along with reference values.
    \begin{ruledtabular}
    \begin{tabular}{cccccc}
         & $\rho (g cm^{-3})$ & K (GPa) & $\langle d_{Zn-Zn}  \rangle$ (\AA) & $\langle d_{Zn-N}\rangle$ (\AA)& $\langle \theta_{N-Zn-N} \rangle$ (\textdegree) \\
         \hline
         nb-ZIF-FF&  1.57& 3.9 & 5.81 & 1.86 & 109\\
         Reference&  1.63\footnotemark[1]& 4.3\footnotemark[2] & 5.95\footnotemark[3] & 1.99\footnotemark[3] & 104\footnotemark[4]\\
    \end{tabular}
    \end{ruledtabular}
    \footnotetext[1]{Ref.~\onlinecite{Bennett2016}.} 
    \footnotetext[2]{Ref.~\onlinecite{Castel2023}.} 
    \footnotetext[3]{Ref.~\onlinecite{Gaillac2017}.} 
    \footnotetext[4]{Ref.~\onlinecite{Beake2013}.} 
\end{table}

Besides correctly capturing average distance and angle values, nb-ZIF-FF also passes an even more difficult test: it yields correct distributions for these properties. To the best of our knowledge, the angles distributions have up to date only been correctly predicted by \textit{ab initio} molecular dynamics, this is the first time that an atomistic force field incorporating reactivity allows to accomplish this challenging task, particularly for the N--Zn--N angle distribution that is known to slightly widen in the amorphous phase.~\cite{Beake2013} Indeed, as pointed out by Castel and Coudert,~\cite{Castel2023} the most widely used reactive force field, ReaX-FF~\cite{Yang2018}, fails to reproduce the N--Zn--N angle distribution. Fig. S1 in the Supplementary Material shows that nb-ZIF-FF is capable of predicting the subtle widening of the angle distribution, that is due to the presence of a small proportion of tri-coordinated Zn cations in ZIF$\_$a. 

\begin{figure}
\includegraphics[width=0.47\textwidth]{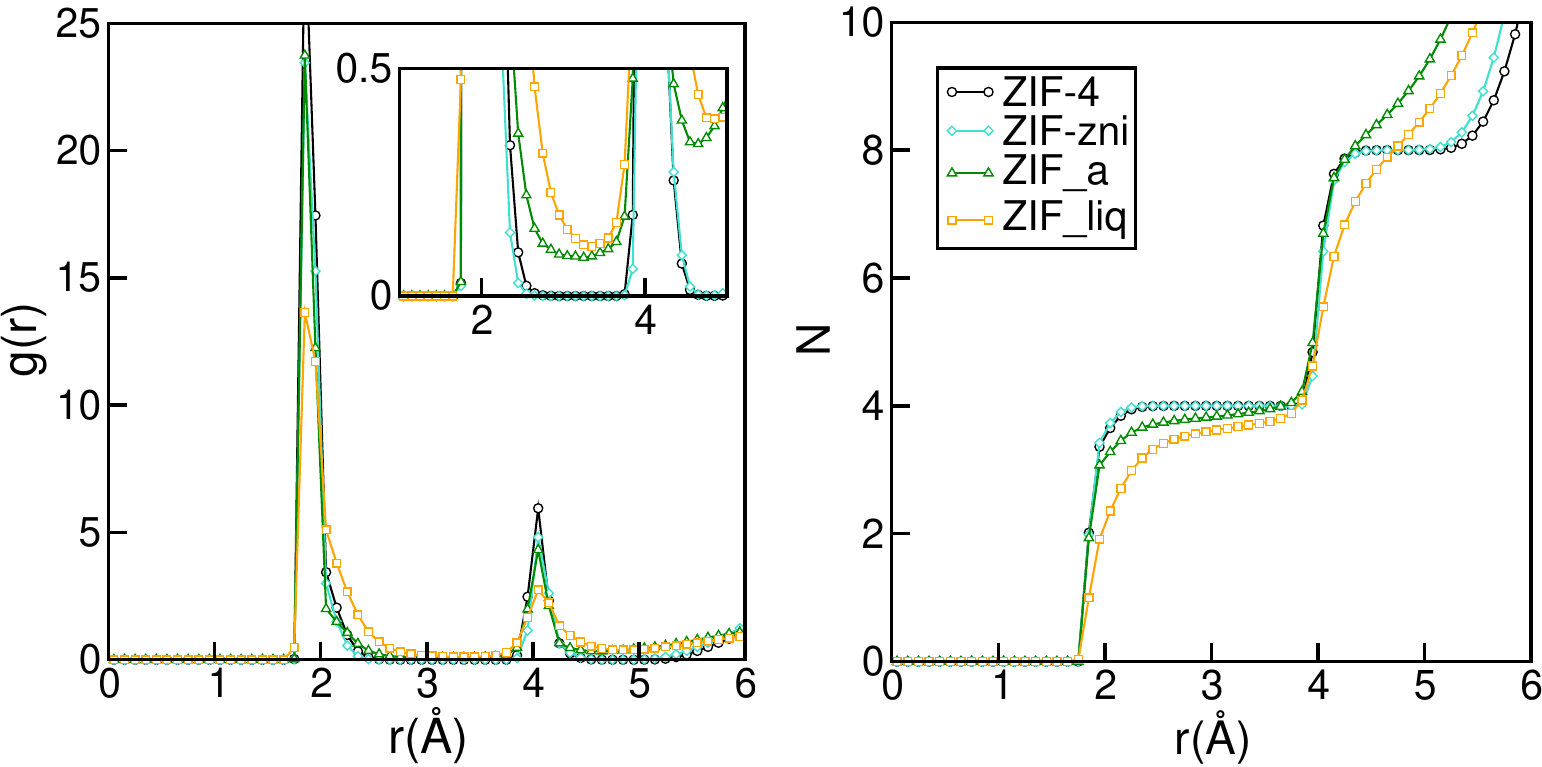}
\caption{\label{fig:rdfs_integral} Zn--N radial distribution (left panel) and its integral (right panel) for ZIF-4, ZIF-zni, ZIF$\_$a and ZIF$\_$liq obtained by molecular dynamics simulations employing nb-ZIF-FF as a force field.}
\end{figure}

The subtle under coordination of the \ce{Zn^2+} centres in the amorphous material can be appreciated in Fig. ~\ref{fig:rdfs_integral}. The left panel shows the Zn--N radial distribution function for the four phases studied. Even though all structures are quite similar and difficult to distinguish experimentally,~\cite{Baxter2015} we can already see some subtle differences in their RDFs. Indeed the peaks for the liquid are broader than those from the amorphous solid and those are, in turn, broader than for the crystals. In the inset of the left panel, we can see that the RDF does no go to zero between the first two peaks nor for ZIF$\_$a nor for ZIF$\_$liq, which indicates nitrogen exchanges in the first coordination sphere of the zinc cations and is in agreement with previous findings.~\cite{Gaillac2018}
The right panel of Fig.~\ref{fig:rdfs_integral} shows the integral of the Zn--N RDF for the four phases, which gives us information on the number of N that can be found on average around a \ce{Zn^2+} centre as a function of the distance. The curves for ZIF-4 and ZIF-zni reach the value of four at 2.3 \AA \  indicating the expected perfect tetrahedral coordination in their first neighbours sphere.
On the other hand, the curve corresponding to ZIF$\_$a shows a loss in the average Zn--N coordination for the same distance, which is even more pronounced in the case of ZIF$\_$liq.
This indicates that both disordered phases typically contain a small proportion of tri-coordinated \ce{Zn^2+} centres, as a result of the increased flexibility of ligand motions, as found experimentally.~\cite{Ryder2017}

The excellent performance of the combination of nb-ZIF-FF with our simulated annealing method in yielding ZIF$\_$a configurations in good agreement with several experimentally measured properties make us confident in our model for this glassy material.

\subsection{\label{sec:descriptors}Machine Learning Methods}

To succeed in our goal of studying the mechanism of the ZIF-4 and ZIF-zni ordered--disordered phase transitions, we first need to define a metric that allows us to distinguish between these crystalline phases and their disordered counterparts at the local, atomic environment level. Such a local metric would allow us to track the evolution of the amorphization and melting processes throughout a simulation, to observe their progression from the deformation of the local structure around a single \ce{Zn^{2+}} centre (i.e. distortions in the tetrahedral spacial arrangement of the first ligand neighbours) to the generation of larger disordered domains within the crystal and how these domains propagate or merge until they take over and the phase transformation reaches its end.

Amorphous materials have traditionally been characterised via en extensive exploration of structural properties, including RDFs, structure factors and pore size distribution.~\cite{Sapnik2023} 
These ways of describing amorphous materials, albeit useful, rely on preconceived ideas of the underlying chemistry of these materials. Machine learning methods based in agnostic descriptors have been developed to automatically identify, differentiate and classify long-range ordered materials in an unbiased way, free of preconceptions. The task becomes more daunting when the collection of objects to be classified includes amorphous materials, due to the inherent lack of translational symmetry which makes both the choice of appropriate descriptors as well as the sampling of structural diversity, harder. A number of methods have been proposed to deal with this difficult task.~\cite{Karayiannis2009, Pietrucci2015,Caro2018,Swanson2020,Banik2023,Rogal2019} Here, we test the performance of Behler-Parrinello Symmetry Functions (BPSF)~\cite{Behler2011} as unbiased generic descriptors that act at the atomic environment level, to inform the degree of ZIF$\_$a-ness, ZIF-4-ness, ZIF-zni-ness and ZIF$\_$liq-ness of a given \ce{Zn^2+}-centred environment. These functions fulfil the basic criteria for being well-behaved chemical descriptors, that is, they yield the same value for two configurations that are related in that one of them is the result of a translation, rotation or same-element-atom permutation operation applied over the other. These atom-centred many-body functions can be classified into two types: \emph{radial} and \emph{angular}. The former ones are given by the sum over two-body terms and are related to the connectivity of the central atom, while in the latter ones, three-body terms are considered. We tested twelve symmetry functions of the type:

\begin{eqnarray}
G_i^{rad}=\sum_{j \neq i} e^{-\eta(R_{ij}-R_{s})^2} \cdot f_c(R_{ij}) 
\label{eq:sfrad}
\end{eqnarray}

\begin{eqnarray}
G_i^{ang}=2^{1-\zeta}\sum_{i,j,k} (1-\lambda \cos{\theta_{ijk}})^\zeta \cdot e^{-\eta(R_{ij}^2+R_{ik}^2+R_{jk}^2)} \nonumber
\end{eqnarray}
\begin{eqnarray}
\cdot f_c(R_{ij}) \cdot f_c(R_{ik}) \cdot f_c(R_{jk})
\label{eq:sfang}
\end{eqnarray}

Where $f_c$ is a cutoff function that decays to zero at a distance $R_c$. Each of the $n_{Zn^{2+}}$ cations in a given configuration will be thus characterised through 12 symmetry functions, each of them centred on the tagged \ce{Zn^{2+}} cation, 4 of them radial and 8 angular. Only Zn--Zn correlations were considered. Further details, including the parameters that were considered for defining the symmetry functions can be found in the Supplementary Material.

Before analysing our trajectories through the lens of our chosen descriptors, we sought to verify their aptitude in recognising the four systems studied. To this end, we built a database comprising 
30000 configurations in total, out of which 3000, 3000, 12000 and 12000 correspond to ZIF-4, ZIF-zni, ZIF$\_$a and ZIF$\_$liq respectively, each of them comprising 64 atomic \ce{Zn^2+}-centred environments. The amount of non-crystalline structures is higher to improve the classification, since the differentiation between these two groups will be the most challenging task for the algorithm. 
The configurations correspond to microstates obtained from MD simulations at temperatures spanning the whole stability range in the case of crystalline structures, while for ZIF$\_$a the sampling was made at temperatures between 300 K and 500 K. As mentioned above, liquid state configurations were sampled at 700 K. We justify this criterion by the change of slope at $\sim 600 K$ observed in the curve of mean potential energy vs. temperature (see Fig. S2) associated with the jump in $Cp$ that occurs at the glass transition temperature (ZIF$\_$a $\xrightarrow{}$ ZIF$\_$liq). 
We then computed the symmetry functions described by eqs.~\ref{eq:sfrad} and ~\ref{eq:sfang} for each of their \ce{Zn^{2+}}-centred environments \textit{via} the RuNNer code~\cite{Behler2015} and plotted their values distributions for all structures together as box plots in R,\cite{rrr} the resulting graph is shown in Fig. S3. We can see from the plot that none of the symmetry functions alone suffices to distinguish between the four phases. We thus used them as features to feed into a neural network that was trained to output the probabilities that the \ce{Zn^2+}-centred environment belongs to a ZIF-4, ZIF$\_$a, ZIF$\_$liq or ZIF-zni phase (four output values). 
In order to train the neural network, we divided our database, composed by 1920000 \ce{Zn^{2+}}-centred environments (64x30000 structures) into train and test sets in a 80:20 proportion.  
The neural network architecture was composed by an input layer of 12 nodes, corresponding to the symmetry functions, a single hidden layer comprising 6 nodes, and a output layer of 4 nodes, each one representing the probability of an environment to be classified as one of the reference structures (see Fig.~\ref{fig:ML}). Further details can be found in the Supplementary Material. If we assign the environment to the class that has the highest associated probability to obtain a clear-cut sorting, our neural network yields 90.3\% accuracy in the classification exercise for the test set.

\begin{figure}
\includegraphics[width=0.48\textwidth]{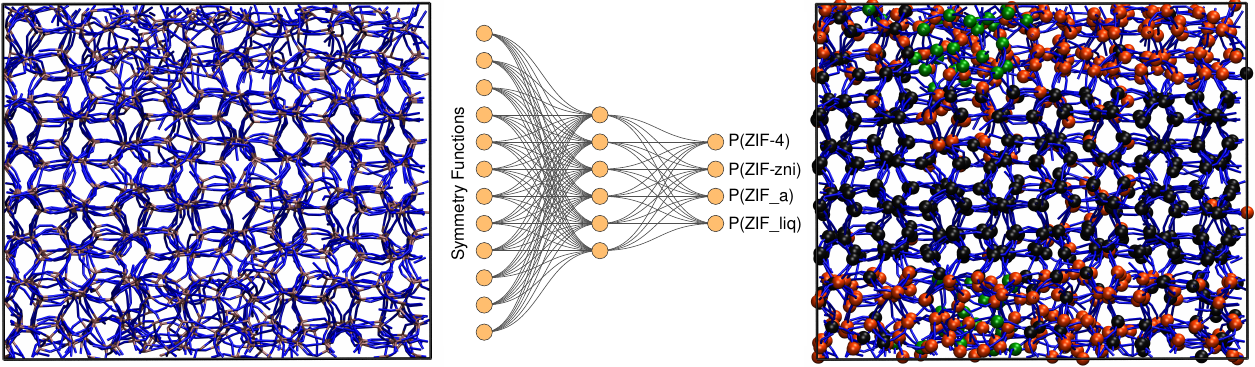}
\caption{\label{fig:ML} Schematics of our neural network trained over atom-centred symmetry functions of ZIF-4, ZIF$\_$a, ZIF-zni and ZIF$\_$liq microstates that allows to classify \ce{Zn^2}-centred environments (left and right hand illustrations: before and after the classification.}
\end{figure}

In table \ref{tab:confusion} we show the confusion matrix obtained for the test set (which is composed of structures that were not used in the training process). This table indicates the fractions of environments of each type (rows) classified as each of the possible structures (columns). The diagonal elements correspond to the fraction of correct classifications.
As expected, the highest source of error ($\sim$10\%) was obtained from the miss-classification of ZIF$\_$a and ZIF$\_$liq structures,~\cite{Schoenholz2016} while the classification of crystalline structures was satisfactory in 95\% of the cases.
Indeed, by simple visual inspection both disordered phases seem to be indistinguishable (see Fig.~\ref{fig:mofs}). A hint that the classification was possible was given by the fact that the Zn--N radial distribution functions for ZIF$\_$a and ZIF$\_$liq differ, as shown above in Fig.~\ref{fig:rdfs_integral}a. It is however quite remarkable that the neural network manages to distinguish them with such low error despite their important structural similarity. The accuracy of the classification is even more impressive if we take into account that it was made based on structural information only, as it is well-known that the main difference between glasses and their parent crystalline structures lie in their dynamics rather than in their structures. Incorporating dynamics information into the features for a neural network distinguishing amorphous materials would most likely further improve the accuracy. 

\begin{table}
\centering
\caption{\label{tab:confusion}}{Confusion Matrix. The diagonal elements are highlighted in bold and the predicted classes in italics.}
\begin{tabular*}{.32\textwidth}{l@{\extracolsep{\fill}}|ccccc}
& \textit{ZIF-4} & \textit{ZIF-zni} & \textit{ZIF$\_$a} & \textit{ZIF$\_$liq} \\
\hline
ZIF-4  & \textbf{0.951}  & 0.000 & 0.004  &  0.045  \\ 
ZIF-zni & 0.000 &  \textbf{0.946} &  0.044 &  0.010  \\ 
ZIF$\_$a   & 0.000 &  0.019 &  \textbf{0.895} &  0.086  \\
ZIF$\_$liq       & 0.012 &  0.002 &  0.098 &  \textbf{0.888}  \\
\end{tabular*}
\end{table}

Fig.~\ref{fig:ML} shows an example of how our neural network can be applied to follow the time evolution of an amorphization process. In the left part, we can see a material that exhibits coexisting ordered and disordered domains. By applying our neural network, we can distinguish between ZIF-4-, ZIF-zni-, ZIF$\_$a- and ZIF$\_$liq-like environments. 
Finally, to check that the classification of the amorphous phases is meaningful, we plotted the fraction of disordered-like centres that are classified as being ZIF$\_$liq-like instead of ZIF$\_$a-like as a function of temperature (see Fig. S4). We can see from this plot that there is a clear temperature dependence on the labelling, which is consistent to what is expected for the thermodynamic stability of these two distinct disordered states (see Fig. S2).

\subsection{Molecular Dynamics Simulations}

Classical molecular dynamics simulations were carried out through the LAMMPS open source simulation package\cite{lammps} with nb-ZIF-FF as a force field.\cite{Balestra2022} 
The integration of equations of motion was performed in the NPT ensemble, with Nose-Hoover thermostats and barostats. The damping parameters were set to 100 time steps for the thermostat and 1000 time steps for the barostat. Unless the contrary is specified, the barostat alters the box sizes in a isotropic way. In all cases the pressure was set to 1 bar. The time step was set to 0.5 fs except for simulations with temperature over 700 K, in which a time step of 0.25 fs was used. 
For generating configurations for the neural network training set, systems of 64 Zn atoms where used, which corresponds to a 2x2x1 supercell of ZIF-4 or a ZIF-zni unit cell (1600 particles in total, including both crystallographic and dummy atoms) while for production runs, the number of Zn atoms was 1024, which corresponds to a 4x4x4 and a 2x2x4 supercell for ZIF-4 and ZIF-zni respectively.

\section{\label{sec:res}RESULTS AND DISCUSSION}

\subsection{Thermodynamic Analyses}

We start our study of ZIF-4 $\xrightarrow{}$ ZIF$\_$a (amorphization, \emph{transition 1}) and ZIF-zni $\xrightarrow{}$ ZIF$\_$liq (melting, \emph{transition 3}) by determining their equilibrium temperatures. We could be tempted to extract the transition temperature from the simulations that we performed to generate the amorphous structures (see Sec.~\ref{sec:amorphgen}) as the temperature that matches the drastic volume change of the system, which signals that the amorphization process has taken place. However, this would be misleading: the temperature ramp is so fast that the system cannot reach thermal equilibrium at each temperature. The false transition temperature we see when we apply a temperature ramp to heat the system has to be higher than the thermodynamic transition temperature. In order to find the equilibrium temperature for \emph{transitions 1} and \emph{3}, we need to find conditions in which both phases are equally stable.~\cite{Govers2008} We first generated a simulation box where ZIF$\_$a and ZIF-4 coexist and occupy approximately the same volume each by running a short simulation at the temperature in which we observed the phase transition in the simulated annealing (T=550 K). Subsequently, we started a series of molecular dynamics simulations at the NPT ensemble, each with different target constant temperature. We let the system evolve, and then we assessed whether the number of amorphous sites had increased, decreased or remained stable. The temperature for which the two phases coexist without one of them gaining terrain over the other is the equilibrium temperature. We followed the same procedure to determine the ZIF-zni melting temperature. Fig. S5 shows the time evolution of the number of liquid-like \ce{Zn^{2+}} centres for ZIF-zni at different temperatures. We can see that the number of liquid-like centres fluctuates around a constant value at T = 625 K. Through this procedure we found $T\_{amorphization}=T_1=420 K$ and $T\_{melting}=T_3=625 K$. The experimental counterparts are around 520 K and 860 K respectively.~\cite{Widmer2019}

We can also obtain the amorphization and melting entropies ($\Delta S_1$ and $\Delta S_3$) from these simulations. At the temperature in which the phases coexist, their chemical potentials are equal and $\Delta G =0$, so the entropy can be readily obtained from the enthalpy by diving it by $T$. The enthalpy is given by the internal energy coming from the force field plus the pressure-volume term. In addition, we can obtain free energy differences at temperature $T$ by integration of the Gibbs-Helmholtz equation:

\begin{equation}
    \Delta G(T) = -T \int_{T_{eq}}^{T} \frac{\langle\Delta H\rangle(T)}{T^2} dT
\end{equation}
Where $T_{eq}$ is the equilibrium temperature obtained before. This allows us to compute $\Delta G$ of both reactions at 298K.
Then, since we have a common reference state, i.e. the glass, we can extract information about the relative stability of ZIF-4 with respect to ZIF-zni. These results, along with reference values, are presented in Tab.~\ref{tab:thermodynamic}. \\

Our computed enthalpies are in good agreement with the reference experimental and \textit{ab initio} values. From the free energy difference we obtain the correct thermodynamic stability trend: ZIF-4 is in fact metastable at ambient temperature and pressure. Entropies can be rationalized in terms of structural properties. Indeed, the disordered phases have higher entropies than the ordered ones, and if we compare the entropies associated to the crystalline phases, we can see that is higher for ZIF-4 than for ZIF-zni. This is a consequence of the lower density (higher porosity) of ZIF-4, which confers it the possibility of adopting many more equivalent microstates than its high density counterpart can. We note that our equilibrium temperatures are deviated from the experimental values. Even though we cannot predict the right values, the tendencies are reproduced. Experimental values should also be carefully considered, since it has been proven that the amorphization temperature is very sensible to the temperature ramp used to trigger it.~\cite{Bennett2015}

\begin{table}
    \caption{\label{tab:thermodynamic}} Thermodynamic properties of the different transitions studied. For ZIF-4 $\xrightarrow{}$ ZIF$\_$a and ZIF-zni $\xrightarrow{}$ ZIF$\_$liq transitions, the values correspond to the calculated equilibrium temperatures, while the ZIF-4 $\xrightarrow{}$ ZIF-zni values correspond to ambient conditions. Reference values from experiments (\footnotemark[1]) and \textit{ab initio} calculations (\footnotemark[2]) are shown between parenthesis. 
    \begin{ruledtabular}
    \begin{tabular*}{0.7\textwidth}{c|cccc}

         & $T$ &  $\Delta G$ & $\Delta H$ & $\Delta S$ \\
        & $(K)$ &$(kJ mol^{-1})$ & $(kJ mol^{-1})$ &$(J K^{-1} mol^{-1})$ \\
         \hline
         ZIF-4 $\xrightarrow{}$ ZIF$\_$a & 420 (589)\footnotemark[1] & 0.0 & 3.6  & 8.7 \\
         ZIF-zni $\xrightarrow{}$ ZIF$\_$liq & 625 (863)\footnotemark[1] & 0.0 & 16.3 (9.8)\footnotemark[1] & 26.0 \\
         ZIF-4 $\xrightarrow{}$ ZIF-zni & 298 & -6.5 & -12.4 (-14.0)\footnotemark[2] & -19.6   \\
    \end{tabular*}
    \end{ruledtabular}
    \footnotetext[1]{Ref.~\onlinecite{Bennett2015}.}
    \footnotetext[2]{Ref.~\onlinecite{Widmer2019}.}
\end{table} 

\subsection{Mechanisms of the phase transitions}

We continue our study by simulating the amorphization and melting processes at the NPT ensemble at P=1 bar and T=550 K for ZIF-4 and T=700 K  for ZIF-zni. These temperatures are higher than the respective transition temperatures, to guarantee thermodynamic feasibility and reasonably fast kinetics. At the end of these simulations, we reach the corresponding disordered phase (glass from ZIF-4 and liquid from ZIF-zni). We also run the same simulations for ZIF-4 and ZIF-zni but with a box consisting of 3x3x3 unit cells to verify whether the size of the simulation box was large enough to avoid unphysical effects that could happen if the process was favoured by an early artificial percolation of the new phase in a small simulation box. The average times needed for the amorphization to occur for the smaller systems are very close to those we obtained for the larger ones, so we can confirm that our boxes are large enough to adequately treat the two processes. We run four independent systems for each transition in order to take into consideration their stochastic character, the results presented below come from an average of these four independent simulation experiments. 

\begin{figure*}
\includegraphics[width=
\textwidth]{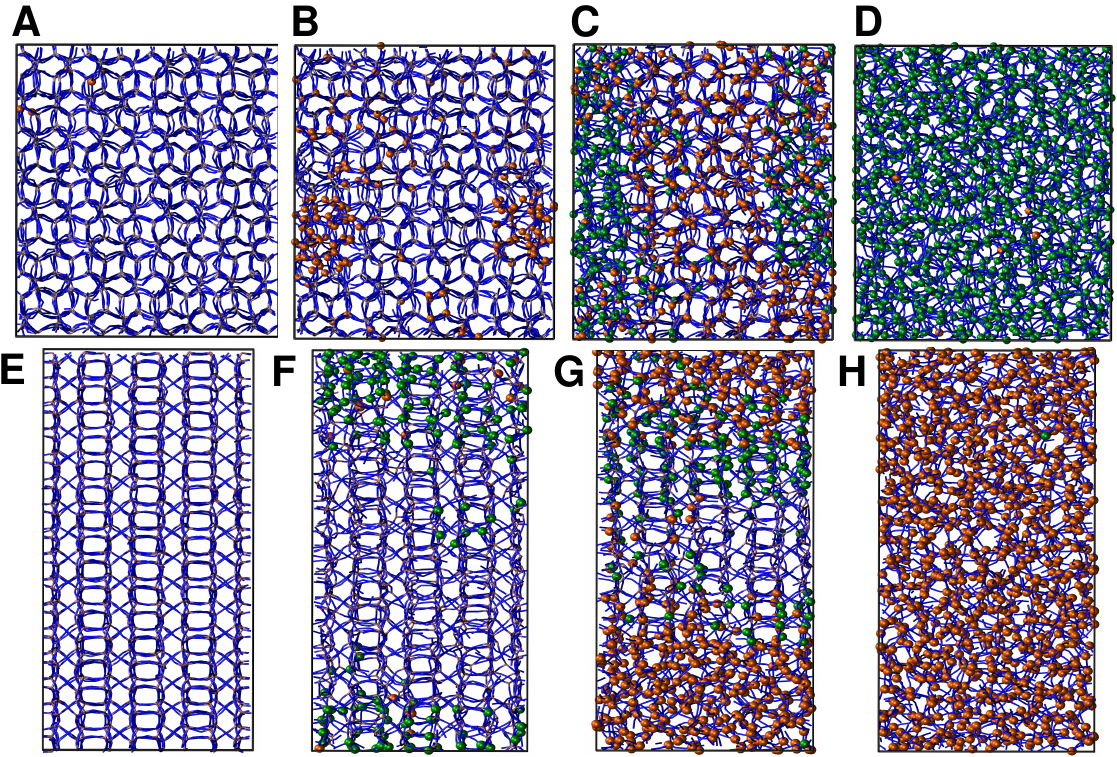}
\caption{\label{fig:zif4time} A--D: Representative snapshots of a ZIF-4 amorphization simulation at (A) 0, (B) 1.1, (C) 1.7 and (D) 4.5 ns. 
E--H: Representative snapshots of a ZIF-zni melting simulation at (E) 0, (F) 0.5, (G) 0.8 and (H) 1.25 ns. 
ZIF$\_$liq- and ZIF$\_$a-like environments are shown in orange and green respectively. For clarity purposes, \ce{Zn^{2+}} centres classified as belonging to a ZIF-4 and ZIF-zni phases are not coloured.}
\end{figure*}

A series of snapshots of the system throughout the simulation is shown in Fig.~\ref{fig:zif4time}. Panel A shows the initial configuration, which corresponds to a ZIF-4 crystal. As expected, and confirming the validity of our neural network, the amount of environments classified as ZIF-zni-like remains negligible during the whole simulation. We first observe the generation of a liquid-like phase (coloured in orange, panel B). Subsequently, the amorphous ZIF$\_$a phase is formed from within this lower density phase (coloured in green, panel C) and gradually expands until it takes over the whole system (panel D, final configuration). The amorphization process thus consists of two steps: a first step in which the Zn--N connectivity slightly drops to yield a more disordered, low density state, followed by a second stage that gives rise to ZIF$\_$a. The intermediate liquid-like phase that is first formed could be associated to the experimentally identified low density non-crystalline phase.~\cite{Bennett2015} Indeed, this phase is less dense than ZIF$\_$a in about $\sim$ 10\% which is comparable to the experimentally obtained density difference between the low density and high density amorphous phases.~\cite{Bennett2015} 
The liquid phase, in turn, is more dense than the porous ZIF-4 crystal. Note that we
did not explicitly include the experimentally found low density phase in the training of the neural network, but the algorithm identified it as a liquid-like phase probably due to the
correlation between local density and the symmetry functions
values. 
Furthermore, since the classification is done only by structural information, the fact that the intermediate phase is assigned to be a liquid does not imply that the dynamics are faster than those associated to the amorphous final phase. 

For ZIF-zni (Fig.~\ref{fig:zif4time}, panels E-H), the situation is inverted: from the crystal (panel E), the connectivity starts gradually diminishing to yield an amorphous-like phase at a first stage (panel F), which then continues to increase its density to reach the final ZIF$\_$liq state (panel G). The liquid phase grows inside the amorphous phase until it takes over the whole simulation box. This suggests that both transitions are thus two-step processes, in what they have in common is that the intermediate phase that is formed in the first step has intermediate density between that of the initial and the final phases ($\rho_{ZIF-4}$ < $\rho_{ZIF\_liq}$ < $\rho_{ZIF\_a}$ < $\rho_{ZIF-zni}$).

In all simulation experiments, we observed that the disordered phases grow as a single cluster that expands and reorganises in terms of the Zn--N connectivity, instead of forming a series of disordered micro-domains that subsequently aggregate. The formation of the higher density amorphous phase does not seem to occur at the interface between crystalline and liquid environments: on the contrary, it is generated at the bulk liquid and propagates until practically all environments are ZIF$\_$a-like. For the melting of ZIF-zni, we also observe that the final phase is formed within the bulk of the intermediate disordered phase.  

\begin{figure}
\includegraphics[width=0.48\textwidth]{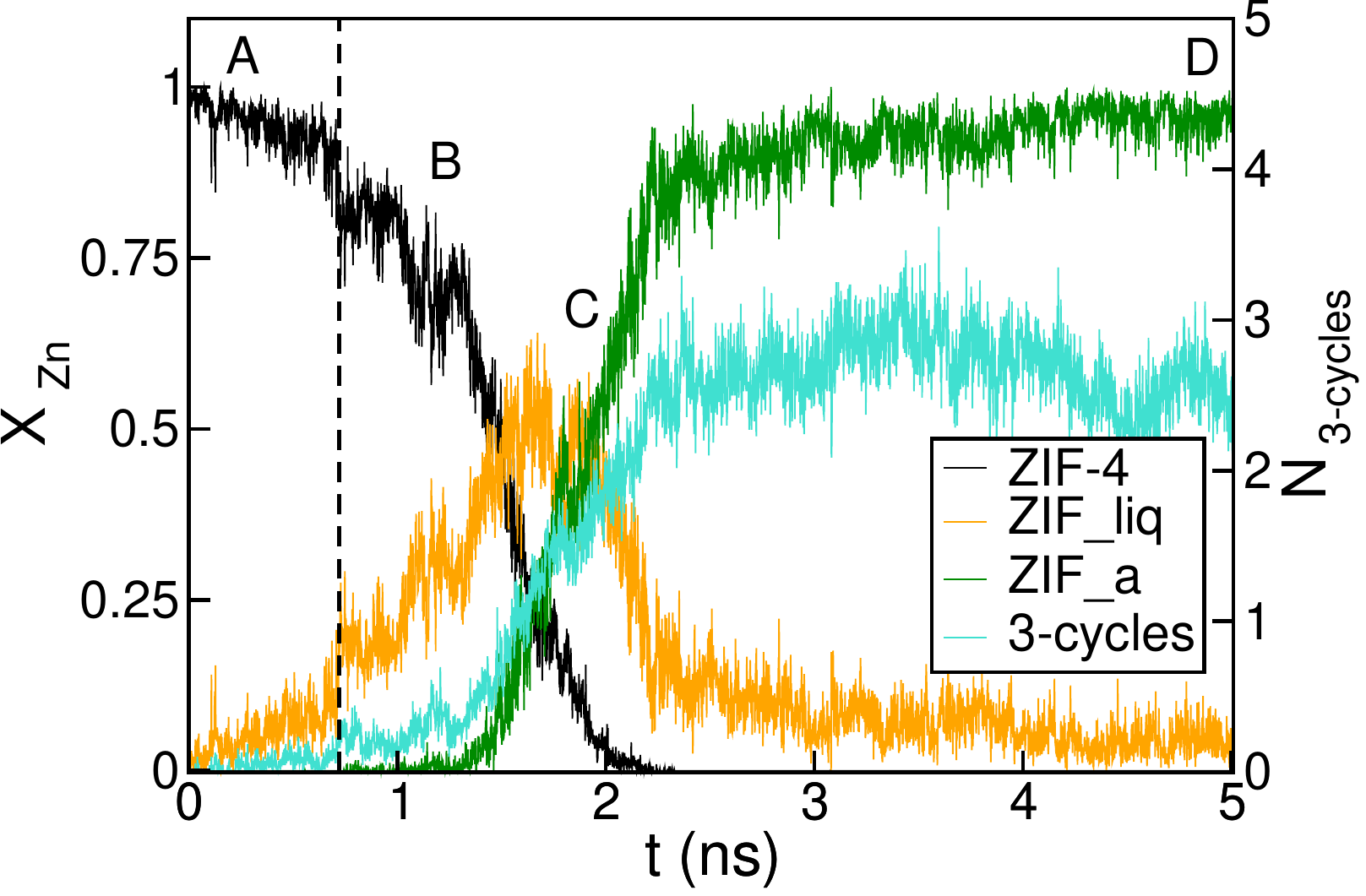}
\caption{\label{fig:zif4clusters} Time evolution of the fraction of disordered (amorphous-, green, and liquid-like, orange) and crystalline (ZIF-4-like, black) \ce{Zn^{2+}} centres for an amorphization simulation starting from a ZIF-4 crystal (ZIF-4 $\xrightarrow{}$ ZIF$\_$a). A,B,C and D mark the points of the trajectory from which we obtained the snapshots for Fig.~\ref{fig:zif4time}. The vertical dotted line indicates the occurrence of a change in connectivity of the type shown in Fig.~\ref{fig:reaction}. The number of 3 membered rings formed by neighbour \ce{Zn^{2+}} centres per unit cell is shown in cyan, the scale is given in the right side \textit{y} axis.}
\end{figure}

The time-evolution of the the fraction of \ce{Zn^{2+}} centres $X_{Zn}$ that correspond to ZIF-4-, amorphous- and liquid-like states is plotted in Fig.~\ref{fig:zif4clusters} and in Fig. S6 for \emph{transition 1} and \emph{3} respectively. 
From these plots we can first confirm the two above mentioned stages for both processes. We can also see in Fig.~\ref{fig:zif4clusters} that once the amorphous cluster is formed, it grows until it takes over the whole simulation box .

To unveil the microscopic processes that trigger the initial steps of the amorphization, we performed a simulation starting from  ZIF-4 at a slightly lower temperature than the above described (T=540 K) in which no amorphization is observed during the whole simulation time span, due to the fact that this is a slow activated process. Indeed, the classification algorithm only identifies a small fraction ($\sim$ 2\%) of non-crystalline \ce{Zn^{2+}} centres, which corresponds to low-lifetime random fluctuations of the network. The presence of tricoordinated \ce{Zn^{2+}} is observed in both cases, and these represent $\sim$ 2\% of the metallic cations. These unstable defects are generated by breaking Zn--N bonds, and have an average lifetime of $\sim$3 ps, meaning that the process is reversible.
This indicates that bond breaking events leading to undercoordinated \ce{Zn^{2+}} centres are not sufficient to trigger the amorphization process. 

\begin{figure}
\includegraphics[width=0.47\textwidth]{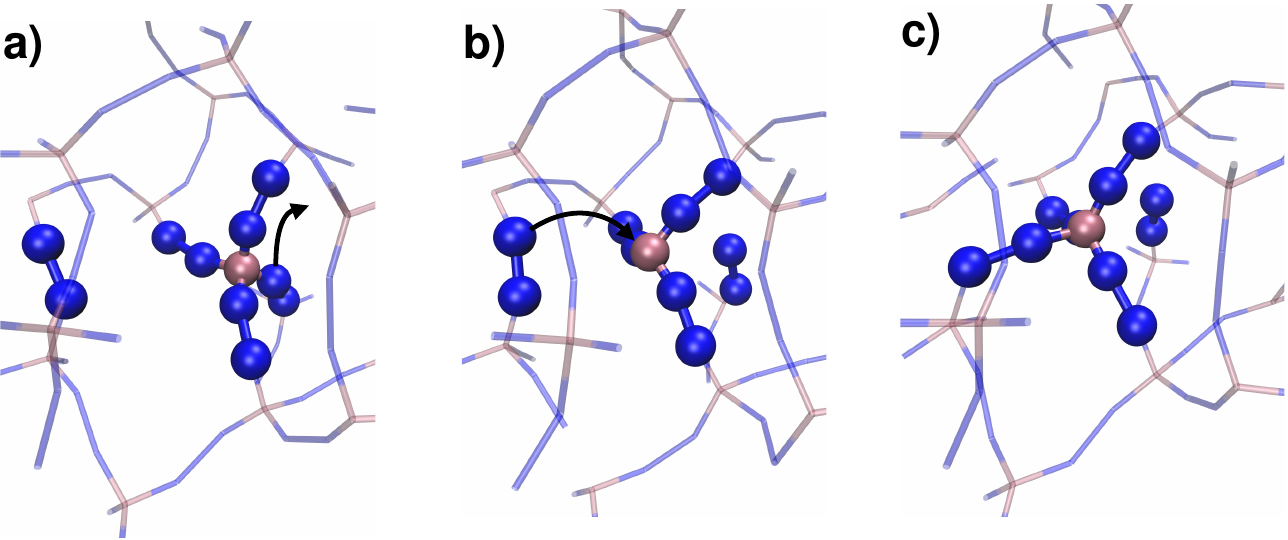}
\caption{\label{fig:reaction} Typical stages of a process of type (iii) described in the main text. a) A Zn--N bond breaks in an initially tatracoordinated \ce{Zn^{2+}}. b) A tricoordinated Zn defect is formed. c) A new metal--ligand bond is formed between the original Zn and a different imidazolate moiety, leading to a change in the connectivity of the system. For clarity purposes only Zn and N atoms are shown in the figure.}
\end{figure}

By monitoring the connectivity of the Zn--N network over time we can observe all possible pathways to restore the tetrahedric coordination of a defect site.
Three different mechanisms can be identified: (i) the broken Zn--N bond is restored, leading to exactly the same connectivity as before the defect formation;
(ii) the tricoordinated \ce{Zn^{2+}} binds to the other nitrogen that belongs to the same ligand moiety, i.e. a rotation of the imidazolate restores the original connectivity; or
(iii) a new bond is formed between the \ce{Zn^{2+}} and a different ligand molecule, leading to a change in the connectivity of the system. This latter mechanism is illustrated in Fig.~\ref{fig:reaction}. The first two mechanisms were observed in both T = 540 K and T = 550 K simulations, but the last one only took place in significant amounts in the high temperature system, which exhibits an amorphization. In the 540 K simulation, just a few (13) events of this kind were registered along the simulation, and every one of them finally reverted to the original connectivity after a short period of time. These may be considered as failed initial attempts of the new phase to nucleate. 
This scenario suggests that the first steps of the amorphization are related to the formation of Zn--N bonds that change the original connectivity of the network. This kind of events are much more infrequent than simple bond breaks or than the formation of tricoordinated sites. Processes of type (iii) typically occur for the first time in a simulation at about t=100 ps, while type (ii) processes rotations take place each 1.5 ps on average and simple bond breaking are observed each 0.1 ps. These results are averages over four independent runs at T=550 K. In figure \ref{fig:zif4clusters} we include a dotted line indicating the time of occurrence of a type (iii) event. We can see that the process is accompanied by a sudden jump in the number of liquid-like environments recognised by the neural network, thus triggering the amorphization transition.

The same analysis was done for the melting of ZIF-zni, by comparing simulations performed at T=700 K and T=670 K, in which no melting was observed, leading to a similar classification of timescales. In this case, since we are at a higher temperature, the characteristic times are reduced to $\sim$50 ps for the first type (iii) event to occur, while type (ii) rotations take place each 0.5 ps on average and bond breakings occur each 0.02 ps.

We complement the microscopic description of the amorphization process by studying the evolution of the number of 3 membered rings formed by neighbour \ce{Zn^{2+}} centres (here we count only the \ce{Zn^{2+}}, in fact the rings are formed by alternating metals and ligands, as the (Si--O)$_{n}$ rings in zeolites). These kind of structures are known to be present in the glass state but not in the crystal, which exhibits rings of 4, 6 and 8 members.\cite{Castel2022}
We can see in Fig.~\ref{fig:zif4clusters} that the number of 3 membered rings in the early stage correlates with the growth of the intermediate low density phase, thus suggesting that the formation of this kind of structures could be an important step in the amorphization process. The correspondence seems to be amplified in the stage of formation of the final glass structure.
This indicates that the final configurations are richer in 3 membered rings patterns, while the intermediate phase preserves more of the topological features of ZIF-4.

\begin{figure}
\includegraphics[width=0.47\textwidth]{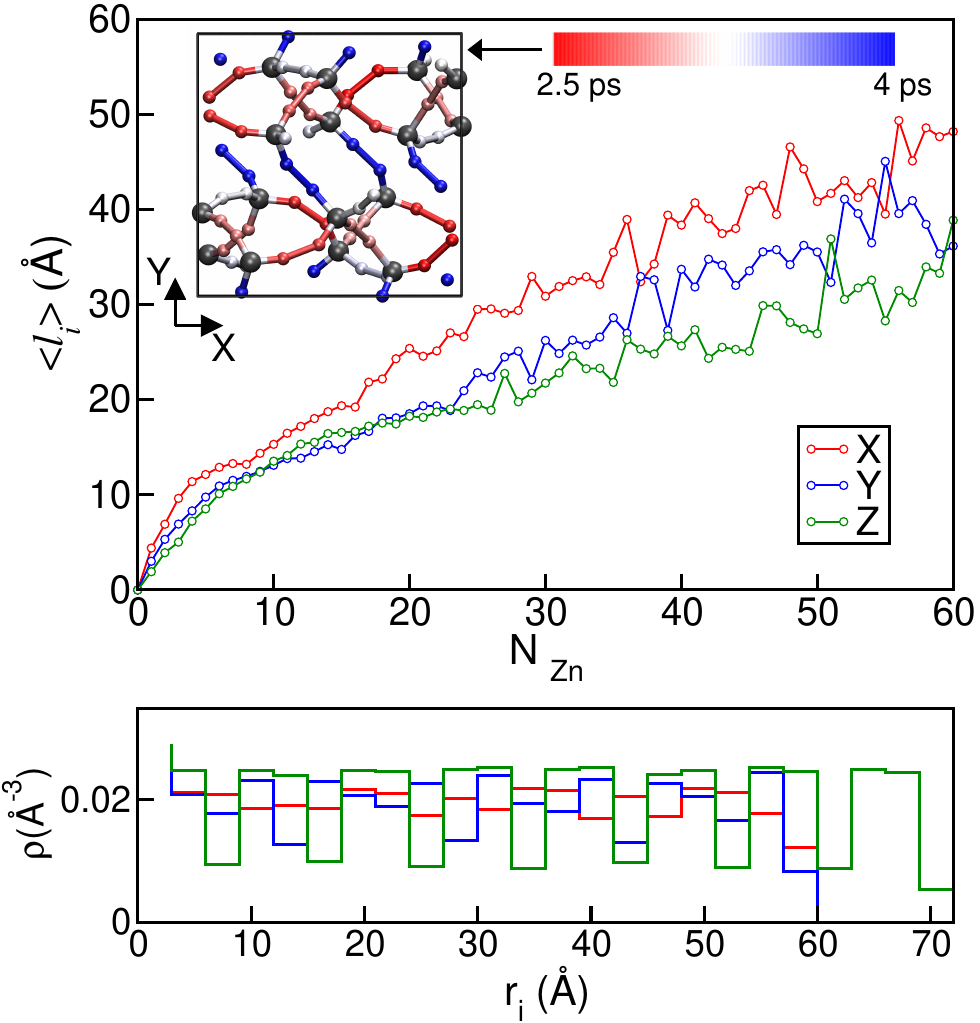}
\caption{\label{fig:aniso} Top: Average length in each of the three cartesian axes directions as a function of $N_{Zn}$, the number of \ce{Zn^{2+}} centres in the ZIF$\_$liq cluster along ZIF-4 amorphization. The inset shows a snapshot of the unit cell in the XY plane. Each ligand is coloured according to its tendency for bond breaking. The colour scale is related to the average time associated to the bond breaking process: from red (less stable bonds) to blue (more stable) passing through white (intermediate reactivity bonds). Only \ce{N} and \ce{Zn^{2+}} atoms (grey) are shown.
Bottom: atomic density as a function of position in each direction for the crystalline ZIF-4 supercell. The colour code is the same as in the top figure.}
\end{figure}

To gain further microscopic insight into the propagation of the disordered phases during these ordered--disordered transitions, we plotted the average disordered (amorphous- and liquid-like) cluster size projected into the three Cartesian axes as a function of the number of disordered \ce{Zn^2+} centres $N_{Zn}$ in Fig.~\ref{fig:aniso}. A similar plot for \emph{transition 3} can be found in Fig. S7. We can see from the curves in Fig.~\ref{fig:aniso} that the growth of the disordered phase is anisotropic: it occurs faster in the $x$ direction than in the other two. This behaviour is reproduced in all the independent simulations we run and can be observed both for the first disordered liquid-like low density phase as well as for the final ZIF$\_$a amorphous phase. In the case of \emph{transition 3}, ZIF-zni melting, no preferential direction can be clearly distinguished during the whole interval of cluster sizes.
   
Anisotropy in the formation of amorphous domains has also been experimentally hypothesised for ZIF-4 systems.~\cite{Widmer2019_2}  
Fig.~\ref{fig:aniso}b depicts local density histograms for ZIF-4 projected into the three Cartesian axes. We can see that the local density stays more or less constant when projected onto the $x$ axis, thus facilitating the connectivity exchange events that drive amorphization, as discussed above. Indeed, the porosity along this direction is more connected than in the other two, which show jumps in the local density that are associated to cage changes. 
We complement this analysis by adding information about the lower temperature simulations made at T=540 K and T=670 K that exhibit bond breaking and formation while preserving the original crystalline structures for ZIF-4 and ZIF-zni respectively. We associated each bond breaking event to a ligand position in the unit cell, in order to check if there are ligand sites that are more labile than others.
We found that in the case of ZIF-4, the 32 ligand sites of the unit cell are divided as follows: 8 of them present a high stability, breaking at a rate of approximately once each 4 ps; 8 of them present the highest tendency to break (once each 2.5 ps), while the others lie in an intermediate reactivity. In the inset of the top panel of Fig.~\ref{fig:aniso} we show a plot of the XY plane of the unit cell with the ligands coloured as blue for the first kind, red for the second, and white for the last one.
We can observe that bonds oriented in the X direction are the most labile, while the ones pointing in Y or Z are more stable, in agreement with our previous results. 
The origin of the difference in stability may be found in the participation of each ligand in different rings. The most stable bonds are those that are part of a four \ce{Zn} membered ring, while the most labile are only part of 6 membered rings.

In the case of ZIF-zni at T= 670 K, the classification of the 128 ligand positions results in 32 members that are more stable than the others, each breaking their coordination bonds each 4.5 ps on average. The others present mean breaking times spanning the interval between 1.9 and 2.7 ps. In this case, members of four \ce{Zn} membered rings present an intermediate stability.
It is found that the most and least labile bonds have important projections in the $z$ direction, this can be observed in the colouring in the inset of Fig. S7. 

\section{\label{Sec:Conc}CONCLUSIONS}

We have studied the molecular mechanisms and thermodynamic properties of a series of phase transformations that link two crystalline polymorph ZIFs (ZIF-4 and ZIF-zni), with two disordered phases, a lower density, liquid-like one (ZIF$\_$liq) and a higher density, amorphous solid (ZIF$\_$a). To this end, we have modelled these states \textit{via} a force field that incorporates reactivity in the coordination bonds. We validate our model by successfully comparing structural and mechanical properties computed from it with reference experimental and \textit{ab initio} data. We have augmented our molecular dynamics method with a phase sorting algorithm based on a neural network that can assign \ce{Zn^{2+}} centred environments to their parent state with high accuracy. The accuracy is quite astounding taking into account that the algorithm has only been fed structural features as inputs and that it can even distinguish between disordered phases. 

Our molecular dynamics simulations allowed us to determine thermodynamic state functions of the ZIF-4 $\xrightarrow{}$ ZIF$\_$a and ZIF-zni $\xrightarrow{}$ ZIF$\_$liq transitions as well as for ZIF$\_$a $\xrightarrow{}$ ZIF$\_$liq \textit{via} thermodynamic integration. Specially relevant are those state functions related to the transition between the two crystalline polymorphs that are difficult to measure experimentally. Our results are in good agreement with reference values, and those that were not previously measured can be rationalized in terms of the disorder and porosity of the different phases.

We have furthermore followed the amorphization of ZIF-4 and the melting of ZIF-zni processes and obtained mechanistic details at the microscopic level. We identified two stages in both processes. Density changes monotonically in both cases, as does the connectivity.  These processes occur \textit{via} changes in the identity of the first neighbour ligand moieties that are coordinatively bonded to the \ce{Zn^{2+}} centres. Kinetics of the amorphization of ZIF-4 are faster in a preferential direction. We have rationalised this through the analysis of local densities projected onto the three Cartesian axes and the study of the lability if the coordination bonds.

This work sheds light on the mechanism of important phase transformations for materials that are excellent candidates for solving pressing environmental and industrial issues. More specifically, our results help better understand the process of generation of amorphous MOFs, which have have been suggested to be key in bridging the gap between research laboratory and real-world application of these promising and versatile porous materials. We also hope that the methods that we developed and deployed are useful to other researchers in this and other fields to study the amorphization of other families of materials or even more broadly to study other kinds of reactive processes. 

\begin{acknowledgments}
This work was funded by the European Union ERC Starting grant MAGNIFY, grant number 101042514. This
work was granted access to the HPC resources of CINES under the allocation A0130911989 made by GENCI and to the HPC resources of the MeSU platform at Sorbonne-Université. 
\end{acknowledgments}

\section*{Data Availability Statement}

The data that support the findings of this study are available within the article and its supplementary material.

\bibliography{1_magnify}

\end{document}